\documentstyle[12pt]{article}
\begin{document}
\begin{center}
{\Large \bf High E$_T$ jets and double scattering}\\

\bigskip
{\large I.M. Dremin} \\

\bigskip
Lebedev Physical Institute, Moscow, Russia
\end{center}
\begin{abstract}
The recently found excess of high E$_T$ jets over current QCD
predictions is attributed to gluon radiation at double scattering
process inside a nucleon. The order of magnitude estimate of the cross
section fits experimental findings rather well. The specific features of
the process are discussed. Various similar hadronic effects related to
the short radiation length are described.
\end{abstract}

It has been recently found [1] that the experimental cross section
for high E$_T$ jets is significantly higher than current QCD
predictions. It is appealing to ascribe this excess to "new" physics, in
particular, to new contact terms in the Lagrangian at the scale of
lengths $10^{-16}-10^{-17}$cm corresponding to transverse jet energies
E$_{T}\geq $200GeV. As the authors of [1] claim, however, such an appeal
"is not defensible" until other traditional mechanisms are fully taken
into account.

We propose the gluon radiation at double scattering process as the
candidate for such a mechanism. It has not yet been taken into account
in current QCD fits. The short radiation length of an impinging quark
suffering two subsequent collisions inside a hadron target is in charge
of high E$_T$ gluon radiation. Here, we provide the order of magnitude
estimates only. The more detailed treatment will be published elsewhere.

First, let us give some qualitative  arguments in favor of it. In the
target rest system, the proton with energy E$_L\sim $1.8 $ \cdot 10^6$GeV
hits the antiproton at rest (E$_{c.m.}\sim $1.8TeV), and one of its
quarks scatters twice on antiquarks located at the distance $l\sim 1$
Fm. The first scattering can trigger the emission of a virtual gluon
which moves, in this system, at the angle of the order of
$10^{-3}-10^{-4}$ to the primary direction of the quark. For the real
particles it would mean the c.m.s. angle close to $\pi$/2, i.e. small
c.m.s. pseudorapidity. Thus, it "separates" from the quark to the
transverse distance ($10^{-3}-10^{-4})l\sim 10^{-16}-10^{-17}$cm
when the quark suffers another scattering. To resolve such a structure,
it should be large transverse momentum $(q_T >200$GeV) scattering. Only
then the gluon can be created as a hadron jet. Such a scattering is a
rare
process, and, consequently, the probability of gluon radiation at high
E$_T$ is strongly suppressed. It can nevertheless become larger than
traditional QCD probabilities which do not take into account the double
scattering radiation.

Electrodynamics analogies are helpful to elucidate the main features,
and they were previously used by the author [2-4].
However, the upper bounds for probabilities were only obtained because
no averaging procedure of the scattering process  was invoked. At high
E$_T$, they strongly overestimate the real probabilities as shown below.
In what follows, for the order of magnitude estimates we shall adopt the
soft processes approximation taking into account the large transverse
momenta in the averaging procedure only. This is justified until
transverse energy becomes high enough to impose the energy conservation
limits.

As well known in electrodynamics [5], the total amplitude for soft
radiation induced by multiple scattering factorizes into the elastic
scattering amplitude and the radiation amplitude. Recently, it was
considered in detail in papers [6-10] with possible extension to QCD.
We adopt here this analogy (see [2-4, 6-8]). The radiation cross section
can be written as
\begin{equation}
\frac {\omega }{\sigma }\frac {d^3 \sigma }{d^3 k}=\frac {\alpha
_{c}C_F} {\pi ^2  \omega ^2}\langle \vert \sum _{i=1}^{N}\vec
{J}_{i}\vert ^{2}+2 Re \sum_{i=1}^{N}\sum_{j=i+1}^{N}\vec{J}_{i}\vec
{J}_{j} (e^{ik(x_{i}-x_{j})}-1)\rangle \label{1}
\end{equation}
for N scattering centers located at points $x_{i}$. The emission current
is
\begin{equation}
\vec{J}_{i}=\frac {\vec{u}_{i}}{u_{i}^{2}}-\frac {\vec{u}_{i-1}}{u_{i-1}^{2}},
\label{2}
\end{equation}
\begin{equation}
\vec{u}_{i}=\frac {\vec{k}_{T}}{\omega }-\frac {\vec{p}_{iT}}{E},
\label{3}
\end{equation}
where $\vec{k}, \omega $ denote the momentum and energy of the emitted
quanta, E is the primary energy, $\vec{p}_{iT}$ is the total transverse
transfer to $i$-th scattering, $\alpha _c$ is strong coupling constant
and $C_{F}=4/3$. The average in (\ref{1}) is done over the transverse
transfers and longitudinal coordinates. This formula is valid in
electrodynamics but does not account for the rescattering of emitted
gluons and the radiation by exchanged gluons in QCD (see [7,8]). It is
used here for qualitative estimates only.

For the double scattering (N=2), one term is only left in the second
part of eq.(\ref{1}). It vanishes at $x_{i}=x_{j}$, i.e. it describes
the radiation emitted in between the two scattering centers. The first
part of (\ref{1}) corresponds to coherent emission on a target as a
single
radiating center and is given by the current QCD fits. We are interested
in the situation when the incoherence due to the interference overwhelms
it. The correction term to the cross section due to incoherent double
scattering is given [3] by
\begin{equation}
\frac {\omega }{\sigma }\frac {d^{3}\sigma }{d^{3}k}=\langle \frac
{4\alpha _{c}C_{F}}{\pi
^{2}\omega ^{2}\theta ^{2}}\sin ^{2}\frac {\omega l\theta ^{2}}{4}
\rangle , \label{4}
\end{equation}
where all the variables are expressed in the laboratory system. The
experimentally measured cross section [1] can be obtained from (\ref{4})
as
\begin{equation}
\frac {d^{2}\sigma }{dE_{T}d\eta }=\langle \frac {8\alpha
_{c}C_{F}\sigma }{\pi E_{T}}\sin ^{2}\frac {E_{T}l\theta }{4}\rangle ,
\label{5}
\end{equation}
where $\eta =-\ln  tg   \theta /2$ is the pseudorapidity.

For E$_{T}\sim $200GeV, $l\sim $1Fm, $\theta <10^{-3}$, the
argument of sin is small. The averaging over the pseudorapidity interval
at a given E$_{T}$ done in [1] leads then to
\begin{equation}
\frac {1}{\Delta \eta }\int \frac {d^{2}\sigma }{dE_{T}d\eta }d\eta
\approx \langle \frac {\alpha _{c}C_{F}\sigma E_{T}l^{2}\theta
_{max}^{2}}{4\pi \Delta \eta }\rangle,  \label{6}
\end{equation}
where $\theta _{max}$ is the l.s. angle corresponding to $\eta =0.1$ in
the c.m.s. It is approximately given by $\theta _{max}\approx
(2m/E_{L})^{1/2}=(4m^{2}/s)^{1/2}$, where  $m$ is a nucleon mass. The
average over longitudinal distances in (\ref{6}) has been done for a
fixed distance $l$ between the scatterers i.e. with $\Delta
(x_{2}-x_{1}-l)$. The averaging prescription of the transverse transfers
asks for special discussion. If one describes the scattering centers by
the screened (at some distance $1/\mu $) Coulomb potential [6,8] then it
means that the average in (\ref{6}) is defined as
\begin{equation}
\langle (\ldots )\rangle =\int \prod _{i=1}^{2} \frac {\mu
^{2}d^{2}q_{iT}}{\pi (q_{iT}^{2}+\mu ^{2})^{2}}(\ldots ). \label{7}
\end{equation}
The integration should start from completely different values of $q_{T}$
for the two centers as discussed above. For soft scattering, it should
begin at $q_{T}=0$ while for hard scattering the low limit is equal to
E$_{T}$. The averages provide, correspondingly, the factors 1 and $\mu
^{2}/E_{T}^{2}$. The final estimate of the cross section reads
\begin{equation}
\frac {1}{\Delta \eta }\int \frac {d^{2}\sigma }{dE_{T}d\eta }d\eta
\approx \frac {\alpha _{c}C_{F}\sigma m^{2}(\mu l)^{2}}{\pi \Delta \eta
s E_{T}}. \label{8}
\end{equation}
It is reasonable to consider $l\sim 2/\mu, \sigma = \sigma _{el}
\approx 10^{-26}cm^{2}, \alpha _{c}C_{F}\sim 0.1$. At E$_{T}$=200GeV,
$(s)^{1/2}=1.8\cdot 10^3$GeV one gets from (\ref{8})
\begin{equation}
\frac {1}{\Delta \eta }\int \frac {d^{2}\sigma }{dE_{T}d\eta }d\eta
\approx 3\cdot 10^{-3}\frac {nb}{GeV}   \label{9}
\end{equation}
to compare with the experimental value [1] (5.11$\pm 0.17)\cdot
10^{-3}\frac {nb}{GeV}$. According to Fig.1 in [1], the current QCD
predictions underestimate the experimental values at E$_{T}\sim
$200GeV by about 10-15$\%$ only. Thus eq. (\ref{8}) overestimates the double
scattering radiation. However, this estimate should only be trusted by
an order of magnitude because of undefiniteness in the choice of $\mu l$
and of the averaging procedure but it shows that the double scattering
process can contribute a noticeable share to the cross section at
E$_{T}$=200GeV.

The energy dependence implied by eq. (\ref{8}) predicts that the similar
effect at energy $(s)^{1/2}$=540GeV could be observed for lower
transverse energies E$_{T}\approx $130GeV. However, the cross section
decrease at larger E$_{T}$ predicted by (\ref{8}) is too slow to fit the
experimental data [1]. There are some reasons for this failure. First,
at larger E$_{T}$ and larger $\theta $ the argument of sin in (\ref{5})
becomes comparable to 1, and it provides additional damping of the cross
section omitted above. Second, the very first scattering can happen also
at larger transfers. The corresponding average can then give rise to a
stronger damping factor decreasing with E$_{T}$. Third, the most
important factor can come out of the decrease of the parton distribution
functions which are not considered here. The more rigorous treatment is
in order to get quantitative results. Nevertheless, the qualitative
estimates are very encouraging. Earlier estimates [2-4] did not take
into account the damping factor $\mu ^{2}/E_{T}^{2}$ due to hard
scattering and gave the upper limits strongly exceeding realistic
values.

At the end, we mention other effects in hadroproduction which
could be related to the gluon radiation from the limited distance. Those
are the peculiar structure of spike centers distribution observed by
NA22 collaboration in pp collisions at E$_{L}$=250-360GeV [11,12] and
the suppression of the accompanying radiation in heavy-quark jets as
compared to light-quark ones in SLD experiments [13] at Z$^0$ peak. The
first effect was theoretically considered in [2-4]. It appears at
smaller E$_{T}$ and smaller energies because $s$ and E$_{T}$ dependence
in eq. (\ref{8}) still favors it to be of comparable size with current
QCD estimates of radiation at a single scattering. The second one is
related to the short lifetime of heavy quarks and is analogous to
earlier proposals [14] to use the specific features of the radiation
from a finite length for measuring the lifetime of the top quark.

Further tests of these effects should be done. For spikes, it must be
E$_{T}$ distributions of spike particles, measured separately for spikes
within and outside the bumps over the background found in [11,12]. There
could be slight excess of higher E$_{T}$ particles in the first group of
events. For the accompanying radiation of heavy quark jets, one should
observe the ring-like structure of it, i.e. the accompanying hadrons are
emitted at comparatively large angles to the direction of flight of
heavy quark.

Altogether, if confirmed, the three effects can provide us the insight
to the nature of QCD radiation at finite length either due to the double
scattering of light quarks or due to the decay of heavy quarks. In its
turn, it would show in more details the internal structure of hadrons.

I am grateful to my colleagues from LPTHE, Univ. Paris-Sud, Orsay for
their hospitality during my stay there when this work was done. The work
is supported by INTAS grant 93-79 and by the Russian Fund for Basic
Research 96-02-16347a.

\newpage
\begin{center}
{\large \bf References}
\end{center}
 \bigskip
1. CDF collaboration, F.Abe et al, Preprint FERMILAB PUB-96/020-E.\\
2. I.M. Dremin, JETP Lett. {\bf 30} (1979) 140; Sov.J.Nucl.Phys. {\bf
33} (1981) 726.\\
3. I.M. Dremin, JETP Lett. {\bf 34} (1981) 594.\\
4. I.M. Dremin, Elem. Part. and At. Nucl. {\bf 18} (1987) 79.\\
5. L.D. Landau and E.M. Lifshitz, {\it Quantum Electrodynamics}, Course
of Theoretical Physics, Vol.4, Pergamon Press, 1978. \\
6. M. Gyulassi and X.-N. Wang, Nucl. Phys. {\bf B420} (1994) 583.\\
7. X.-N. Wang, M. Gyulassi and M. Plumer, Phys. Rev. {\bf D51} (1995)
3236.\\
8. R. Baier, Yu.L. Dokshitzer, S. Peigne and D. Schiff, Phys. Lett. {\bf
B345} (1995) 277.\\
9. R. Baier, Yu.L. Dokshitzer, A.H. Mueller, S. Peigne and D. Schiff,
preprint CUTP 724, LPTHE-Orsay 95-84, BI-TP 95-40.\\
10. R. Blankenbecler and S.D. Drell, preprints SLAC-PUB-95-6875; 6944.\\
11. I.M. Dremin et al, Sov.J.Nucl.Phys. {\bf 52} (1990) 536.\\
12. NA22 Collaboration, N.M. Agababyan et al, Preprint HEN-380 (1996).\\
13. B.A. Schumm, Yu.L. Dokshitzer, V.A. Khoze and D.S. Koetke, Phys. Rev.
Lett. {\bf 69} (1992) 3025.\\
14. I.M. Dremin, M.T. Nazirov and V.A. Saakian, Sov.J.Nucl.Phys. {\bf
42} (1985) 1010.

\end{document}